\documentclass[12pt]{iopart}

\usepackage{graphicx}
\usepackage{epsfig}
\usepackage[utf8]{inputenc}
\usepackage[T1]{fontenc}

\usepackage{hyperref}
\usepackage{units}

\newcommand{\zerodel}{.\kern-\nulldelimiterspace}

\begin{document}

\title{Quantum tight-binding chains with dissipative coupling}

\author{D. Mogilevtsev}
\address{Centro de Ci\^encias Naturais e Humanas,
Universidade Federal do ABC, Santo Andr\'e,  SP, 09210-170 Brazil;
\\
 Institute of Physics, Belarus National Academy of Sciences,
Nezavisimosty Ave. 68, Minsk 220072 Belarus}

\author{G. Ya. Slepyan}
\address{Department of Physical Electronics, School of Electrical
Engineering, Faculty of Engineering, Tel Aviv University, Tel Aviv
69978, Israel}

\author{E. Garusov}
\address{Institute of Physics, Belarus National Academy of
Sciences, Nezavisimosty Ave. 68, Minsk 220072 Belarus}

\author{S. Ya. Kilin}
\address{Institute of Physics, Belarus National Academy of
Sciences, Nezavisimosty Ave. 68, Minsk 220072 Belarus}

\author{N. Korolkova}
\address{ School of Physics and Astronomy, University of St Andrews,
North Haugh, St Andrews KY16 9SS, UK.}

\begin{abstract}

We present a one-dimensional tight-binding chain of two-level systems coupled only
through common dissipative Markovian
reservoirs. This quantum chain can demonstrate anomalous thermodynamic behavior contradicting Fourier law. Population dynamics of individual systems of the chain is polynomial with the
order determined by the initial state of the chain. The chain can
simulate classically hard problems, such as multi-dimensional random walks.

\end{abstract}


 \maketitle

\section{Introduction}

Coherent chains of interacting quantum
systems represent a wide class of physical objects that define the
behavior of matter under different physical conditions. Study of
theoretical models for these objects, which was started by the
epoch-making works by Hubbard \cite{hubbard} and Ising
\cite{ising}, covers both new forms of matter and new types of
interactions \cite{cohen}. One can coherently chain cooled atoms
in optical lattices \cite{cohen}, Josephson qubits in microwave
transmission lines \cite{josephson}, semiconductor quantum dots,
etc. Interactions between systems can be of quite different
physical nature: spin-exchange  interactions or pseudospin
interactions corresponding to the dipole optical transitions
\cite{scully}, tunneling \cite{zoller}, dipole-dipole interactions
\cite{scully2009,scully2010}, photonic interactions
(Jaynes-Cummings-Hubbard model \cite{blatter}) and many others.
Recent applications of these models  led to a
number of new fundamental results. For example, 1D -
chain of tunnel-coupled systems with two ends connected to heat
reservoirs served as a model to justify the Fourier heat
conduction law from the first principles and to provide
microscopic definition of temperature
\cite{gemmer,ventra}. Other results to be
mentioned are the directivity of collective spontaneous emission
\cite{scully2009,confirmation}), possibility to transfer quantum states \cite{sbose}, the spatial propagation of Rabi oscillations
(Rabi-waves) \cite{slepyan2010,slepyan2012}, and quantum optical
nonreciprocity of the medium in timed Dicke state
\cite{slepyan2013}.

It is of particular importance to study new types of
interactions that determine the coherent behavior of coupled
systems. Here we suggest to couple them in a chain by connecting
them pair-wisely to common dissipative reservoirs. It is already
well-known that coupling several quantum systems to the same
dissipative reservoir allows to obtain a number of highly
non-trivial effects. First of all, such a coupling can
create a decoherence-free subspace \cite{lidar,kwiat}. An arbitrary initial state
is eventually transferred into this subspace. Thus, it is possible to produce
non-classical and entangled states via dissipative dynamics even
without any direct interaction between quantum systems
\cite{brawn}. This effect has many potential applications. For
example, it was recently used in practice to protect the
highly-entangled initial polarization states of photons from
dephasing in optical fibers \cite{breuer2014}. Moreover, combining
coupling to the same reservoir with the
nonlinear interaction between quantum systems, it is possible to
produce nonlinear loss generating robustly a wide variety of
non-classical states \cite{SM,SM1,SM2}. By adding
external driving, a generated non-classical
state can be preserved in the presence of an arbitrarily large linear loss \cite{DM2013}.

\begin{figure}[htb]

\includegraphics[width=\textwidth]{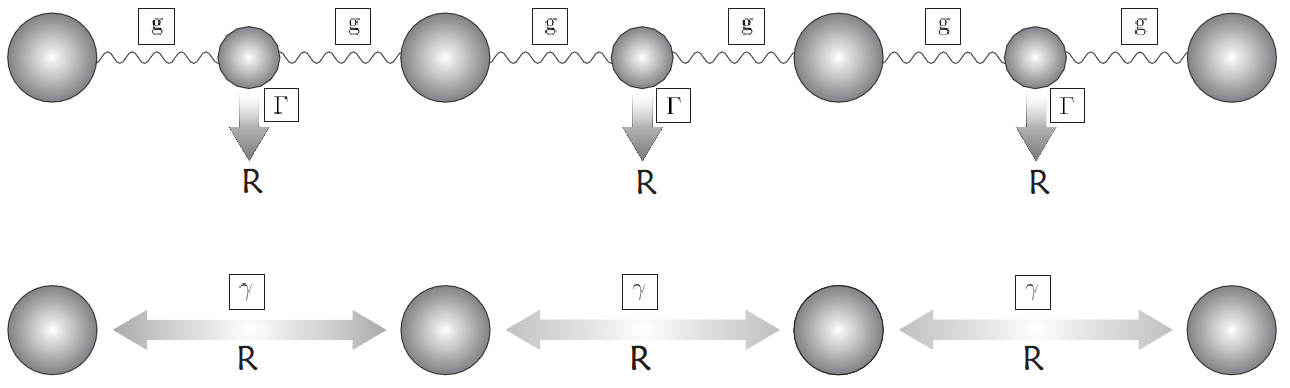}

\caption{(above) A scheme of bipartite tight-binding spin-chain with some elements subjected to strong loss. This scheme is
described by Eq.~(\ref{chain0}). (below) A scheme of dissipatively
coupled tight-binding spin chain described by Eq.~(\ref{chain1}).}
\label{fig1}
\end{figure}

Here we show that by coupling a set of quantum systems through
common Markovian reservoirs and forming one-dimensional
tight-binding chain, it is possible to produce highly non-trivial
dynamics. Such a dissipatively coupled set of just a few two-level
system renders a possibility to reproduce dynamics of far more
complex systems, such as classical heat reservoirs and
multidimensional random networks. This dynamics can be "tuned" by
choice of the initial state of the system.

We demonstrate that even for a few systems in the chain certain groups of matrix elements of the chain density matrix evolve according to equations
formally coinciding with the equations describing classical random walks. For no more than one initial excitation
in the chain, matrix elements of the single-excitation subspace evolve according to the equation describing two-dimensional classical random walk.
Simultaneously, matrix elements describing coherences between the single-excitation and zero-excitation subspaces evolve according to the equation
describing one-dimensional random walk. Taking more than one initial excitation in the chain, one is able to obtain dynamics governed by equations describing multi-dimensional classical random walks. This fact leads to a number of quite-counterintuitive consequences. Notwithstanding the Markovianity of reservoirs, a
decay of excitation in any system of the chain is polynomial with
the same power. By choosing the initial states,  it is
possible to obtain a population decay law $1/t^{2m+1}$, for an arbitrary
$m>0$, provided that the chain
is long enough.   The sum of density matrix elements described by the random walk equation is conserved. So, one can observe rather peculiar thermodynamics-like behaviour of the chain. Despite seemingly classical character, this "thermodynamics" is quite anomalous. The stationary state of the chain can be entangled. Moreover, energy flow through the chain is not described by the Fourier law, whereas the flow of coherences is governed by it.

The outline of paper is as follows. In the Section \ref{secchain} we introduce the concept of tight-binding dissipatively coupled quantum chain. Also, we show how this "quantum gadget" can be built in practice from usual tight-binding unitary chain by selectively applying strong dissipation to certain systems in the chain. In the Section \ref{single} we derive equations for the simplest case of the chain dynamics corresponding to an initial state with no more than just one excitation. Already in this case the chain exhibits a rich variety of highly unusual phenomena. We highlight the thermodynamical analogies in the Section \ref{therm}.
In the Section \ref{polynomial} we analyze in more details possibilities of having different polynomial decay for specific choices of initial states of the chain. Then, in the Section \ref{multidy} we discuss the dynamics of the chain in the case of multiple initial excitations demonstrating how conservation of coherences might arise in this case, too.

\section{The chain}
\label{secchain}

Here we are discussing dynamics of a set of quantum objects
coupled only through common dissipative reservoirs. We are
interested in dynamics arising for the case when our set is rather
large (with the number of objects, $N\gg 1$). In this paper we
consider one of the simplest cases of such a set, namely,  a
tight-binding one-dimensional chain of identical two-level systems
(TLS), e.~g., spins or pseudo-spins. The general scheme of such an
arrangement is as follows.  Each spin is supposed to be coupled to
common Markovian reservoirs with one of two neighbors: $j$th TLS
and $(j+1)$th TLS are coupled to common $j$th reservoir, $(j-1)$th
TLS and $j$th TLS are coupled to $(j-1)$th reservoir, etc.
Reservoirs are supposed to be independent (see Fig.\ref{fig1}
(below)). Thus, we are considering here dynamics of the compound
object described by the following generic effective master
equation:
\begin{equation}
\frac{d}{dt}\rho=\sum\limits_{j=1}^N\gamma_{j}\left(2S_j^-\rho S_j^+-\rho
S_j^+S_j^--S_j^+S_j^-\rho\right),
 \label{chain1}
\end{equation}
where Linbdlad operators are
\[S_j^-=\sigma_j^--\sigma_{j+1}^-\]
and $\sigma_j^{\pm}=|\pm_j\rangle\langle\mp_j|$ are rasing/lowering
operators for the $j$th TLS. Vectors $|\pm_j\rangle$ describe
excited/ground states of the $j$th TLS. Quantities $\gamma_j$ are
relaxation rates into corresponding reservoirs. For simplicity sake,
we consider the finite-size homogeneous chain, thus we take
\begin{equation}
\gamma_{j}=\left\{\begin{array}{ll}
\gamma,& 1\leq j\leq N, \\
0, & j\leq 0,j\geq N+1.
 \end{array}
 \right\zerodel\vphantom{\}}
\label{defgamma}
\end{equation}
where $\gamma$ is a given constant and  $N+1$ is the total number
of TLS in the chain, whereas $N$ is the total number of
reservoirs. The definition (\ref{defgamma}) corresponds to the
total insulation of the chain from the outside areas of space,
i.e., the absence of quantum flux over the chain boundaries.

Notice that Eq.(\ref{chain1}) is the quite general and describes a
1D set of pairwise dissipatively coupled systems. A number of
different physical models can lead to such an equation. For
example, similar band structure of dissipators can arise in a
graphenelike models based on a honeycomb lattice with
nearest-neighbor and next nearest- neighbor couplings (Haldane
model) \cite{mardel1}. Now let us discuss some simple cases as to
demonstrate feasibility of dissipatively coupled chain and
possibility to realize it in practice.

First of all, let us point out that the dissipatively coupled
chain can be produced just by subjecting some elements of a usual
tight-binding chain with exchange interaction (like dipole-dipole
or spin-spin interactions) to strong Markovian loss. Such a
lattice  with the regular pattern of dissipative sites (the every
second one) were considered recently in the context of quantum
walks  and was shown to exhibit such non-trivial effects as
topological transitions \cite{narimanov}. So, let us consider a
simple example of a tight-binding 1D chain of TLS with every
second TLS coupled to the separate bosonic reservoir (see
Fig.\ref{fig1}(above)). Such a system is described by the
following Hamiltonian
\begin{eqnarray}
H_{total}(t)=H_0+H_{chain}+H_{reservoirs}+V_{reservoirs},
 \label{ham0}
\end{eqnarray}
where the Hamiltonian $H_0$ describes non-interacting chain
systems with the transition frequency $\omega_0$ (we are using the
system units with $\hbar\equiv1$),
\begin{eqnarray}
H_{0}=\omega_0\sum\limits_l\sigma_{l}^+\sigma_{l}^-,
 \label{ham00}
\end{eqnarray}
and the Hamiltonian of direct spin-spin  (or dipole-dipole)
interaction is
\begin{equation}H_{chain}=ig\sum\limits_{l}(\sigma_{l+1}^+\sigma_{l}^--\sigma_{l}^+\sigma_{l+1}^-).
 \label{exchange}
\end{equation}
The Hamiltonian, $H_{reservoirs}$, describes modes of independent
bosonic reservoirs coupled to the corresponding TLS of the chain
\begin{equation}H_{reservoirs}=\sum\limits_{lk}\omega_{lk}b_{lk}^{\dagger}b_{lk},
 \label{hreservoir}
\end{equation}
where $\omega_{lk}$ is the frequency of the bosonic mode described
by the annihilation operator, $b_{lk}$, and the creation operator
$b_{lk}^{\dagger}$. The operator $V_{reservoirs}$ describes
coupling of the chain systems to dissipative reservoirs, and can
be represented as
\begin{eqnarray}
V_{reservoirs}=\sum\limits_{lk}(g_{lk}\sigma_{l}^+b_{lk}+h.c.),
 \label{vdiss}
\end{eqnarray}
where  $g_{lk}$ is the interaction constant for coupling of the
$l$-th chain systems with the mode $b_{lk}$.

In the interaction picture with respect to the bosonic reservoirs
Hamiltonian, $H_{reservoirs}$, and the chain Hamiltonian, $H_{0}$,
the total Hamiltonian (\ref{ham0}) becomes
\begin{eqnarray}
H_{total}(t)=H_{chain}+V(t),
 \label{ham000}
\end{eqnarray}
where
\begin{eqnarray}
V(t)=\sum\limits_{l}(\sigma_{l}^+R_{l}(t)+h.c.),
 \label{vdiss1}
\end{eqnarray}
and reservoir operators are
\begin{eqnarray}
R_{l}(t)=\sum\limits_{k}v_{lk}b_{lk}\exp\{-i(\omega_{lk}-\omega_0)t\}.
 \label{roper}
\end{eqnarray}

We assume that our bosonic reservoirs are Markovian, mutually
independent, and initially in the vacuum state. So, the following
relations hold
\begin{eqnarray}
\nonumber
\langle R_l(t)R^{\dagger}_k(\tau)\rangle\approx \delta_{lk}\Gamma_l \delta(t-\tau),\\
\nonumber
\langle R_l(t)R_k(\tau)\rangle=
\langle R_l^{\dagger}(t)R_k^{\dagger}(\tau)\rangle=0,
\end{eqnarray}
where $\Gamma_l$ is the decay rate into $l$-th reservoir,
$\delta(t-\tau)$ is the Dirac delta-function and the averaging
denoted by $\langle \ldots\rangle$ is carried over the states of
reservoirs.

Now let us suggest that only the every second system of the
reservoir is subjected to loss, i.e. $\Gamma_{2l+1}\equiv 0$.
Also, we assume that losses are occurring on the time-scale much
shorter than the dynamics of the excitation exchange described by
the Hamiltonian (\ref{exchange}). Then, taking for simplicity
equal decay rates, $\Gamma_l\equiv\Gamma$,  in the basis rotating
with the frequency $\omega_0$, one arrives at the following master
equation
\begin{eqnarray}
\frac{d}{dt}\bar{\rho}=-i[H_{chain},\rho]+
\sum\limits_{l}\Gamma\left(2\sigma_{2l}^-\bar{\rho}
\sigma_{2l}^+-\bar{\rho}
\sigma_{2l}^+\sigma_{2l}^--\sigma_{2l}^+\sigma_{2l}^-\bar{\rho}\right),
 \label{chain0}
\end{eqnarray}

Let us assume that TLS corresponding to dissipative sites are
initially in the ground states. If the dissipation rate, $\Gamma$,
is high enough in comparison with the strengths of direct
interaction between neighboring spins, $g$, variables
corresponding to TLS with even numbers can be adiabatically
eliminated.  Indeed,  in this case for the dissipative sites we
have
\begin{eqnarray}
\nonumber \langle \sigma_{2l}^-(t)\rangle=0, \quad \langle
\sigma_{2l}^-(t)\sigma_{2m}^-(\tau)\rangle=\langle
\sigma_{2l}^+(t)\sigma_{2m}^-(\tau)\rangle=0,\\
\nonumber \langle
\sigma_{2l}^-(t)\sigma_{2m}^+(\tau)\rangle\approx
\delta_{lm}\exp\{-2\Gamma(t-\tau)\}.
\end{eqnarray}
Then, taking TLS in dissipative sites as reservoirs and deriving
the master equation up to the second order with respect to the
ratio $g^2/\Gamma$ (see, for example, \cite{breuer}), one can
obtain from the master equation (\ref{chain0})
the master equation Eq.(\ref{chain1}) describing dissipatively coupled tight-binding chain
with the dissipative rate  $\gamma\approx g^2/2\Gamma$.

Thus, we have demonstrated one of the ways to obtain the
dissipatively coupled chain. One can get it as the limiting case
of a usual tight-binding chain with some sites subjected to
losses. Such a bipartite dissipative lattice shown in
Fig.\ref{fig1}(b) can be realized in practice in a lot of
different ways. For example, it can be build using a chain of
color-center defects in diamond microcrystallites
\cite{wrachtrup}. For color-center defects in diamond
microcrystallites it is typical to have very low decoherence rates
even at room temperature. Such objects can be manipulated with
high precision and addressed individually by applied external
electromagnetic fields \cite{rittweger}.   Also, chains of
individually manipulated atoms deposited on the metallic surface
can be used for the purpose \cite{khaj1,khaj2,khaj3}. Among
possible perspective candidates one can mentions schemes  with
trapped ions in optical lattices \cite{fridenauer,fridenauer1},
photonic structures described by Jaynes-Cummings-Hubbard model
\cite{blatter},  coupled system of optical waveguides
\cite{white2011}, Bose-Einstein condensates in multiple-well
potentials \cite{SM}, or even networks of pigments in
light-harvesting molecules \cite{pygments,pygments1}.

To realize the chain it is not necessary to manipulate "chain
links" with high precision and address them individually. In our
example, one can avoid coupling strongly dissipative reservoirs to
the every second TLS in the chain. Restrictions imposed on
precision of individual addressing can be significantly relaxed,
if one considers a strongly coupled sufficiently long sub-chain
instead of just one TLS  as a lossy system  to be adiabatically
eliminated (an example of realistic consideration of such a
dissipative sub-chain one can see, for example, in Ref.
\cite{lovett}).

\begin{figure}[htb]

\includegraphics[width=\textwidth]{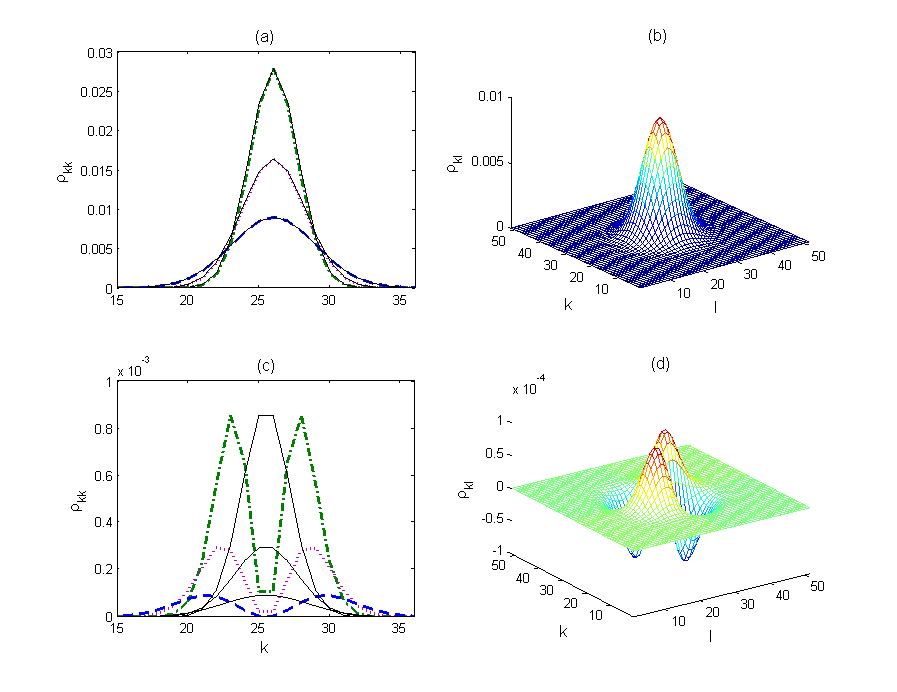}

\caption{(a) A solution of the master equation Eq.~(\ref{chain2}) for
diagonal elements, $\rho_{kk}$, for only one TLS initially excited
(the one in the middle of the chain, $j=26$). Dash-dotted, dotted
and dashed curves correspond to the times $\gamma t=3,5,9$. Thin solid
lines depict results Eq.~(\ref{heat solution1}) given by the heat equation. (b) A
solution of Eq.~(\ref{chain2}) for all elements, $\rho_{kl}$, for
only one TLS initially excited ($j=26$); $\gamma t=9$. (c)
Solutions of Eq.(\ref{chain2}) for diagonal elements, $\rho_{kk}$,
for completely entangled initial states $\rho_{\pm}$ (\ref{initial
entangled}). Dash-dotted, dotted and dashed curves correspond to the
times $\gamma t=3,5,9$ for the initial state $\rho_{-}$, $j=25$.
Thin solid lines correspond to the same times for the initial
state $\rho_{+}$. For comparison, solutions are normalized. (d) A
solution of Eq.~(\ref{chain2}) for all elements, $\rho_{kl}$, for
the initial state $\rho_{+}$, $j=25$; $\gamma t=9$. For all
examples the number of TLS in the chain is $N=51$. } \label{fig2}
\end{figure}

\section{Single-excitation equations}
\label{single}

To clarify essential
features of dynamics prescribed by the master equation
(\ref{chain1}), let us consider the initial chain state confined to the single-excitation subspace.
Then, as follows from Eq.(\ref{chain1}), the single-excitation matrix elements,
\[\rho_{kl}=\langle 1_k|\rho|1_l\rangle,\]
satisfy the following
equation:
\begin{eqnarray}
\nonumber
\frac{d}{dt}\rho_{kl}=-(\gamma_k+\gamma_{k-1}+\gamma_{l}+\gamma_{l-1})\rho_{kl}+\\
\qquad \quad \gamma_k\rho_{k+1,l}+\gamma_{k-1}\rho_{k-1,l}+\gamma_l\rho_{k,l+1}+\gamma_{l-1}\rho_{k,l-1},
 \label{chain2}
\end{eqnarray}
where the relaxation rates, $\gamma_k$, are defined in accordance with conditions (\ref{defgamma}).
Away from the edges, Eq.~(\ref{chain2}) coincides with a standard equation for a
discrete classical random walk in two dimensions in continuous
time, and describes diffusive propagation of the excitation through the chain mediated by emission to reservoirs and re-absorption from them \cite{kempe}.

As follows from Eq.(\ref{chain1}), the coherences between the single-excitation
subspace and the vacuum
\[\rho_{k0}=\langle 1_k|\rho|0\rangle, \quad |0\rangle=\prod\limits_{k=1}^{N+1}|-_k\rangle\]
satisfy the standard classical equation for the one-dimensional discrete classical random walk:
\begin{eqnarray}
\frac{d}{dt}\rho_{k0}=-(\gamma_k+\gamma_{k-1})\rho_{k0}+
\gamma_k\rho_{k+1,0}+\gamma_{k-1}\rho_{k-1,0}.
 \label{cohr1}
\end{eqnarray}

The immediate consequence of the classical form of both Eqs. (\ref{chain2}) and (\ref{cohr1}) is the conservation of sums of corresponding matrix elements:
\begin{eqnarray}
\nonumber
W=\sum\limits_{k,l=1}^{N+1}\rho_{kl}(t)=\sum\limits_{k,l=1}^{N+1}\rho_{kl}(0), \\
F=\sum\limits_{k=1}^{N+1}\rho_{k0}(t)=\sum\limits_{k=1}^{N+1}\rho_{k0}(0).
\label{conservation}
\end{eqnarray}
Eqs.(\ref{conservation}) are a consequence of collective character of coupling to reservoirs. The
chain described by the master equation (\ref{chain1}) has another pure stationary state in addition to the usual lowest-energy one. This state satisfies the equation $S_j^-|\Psi\rangle=0$, for all $j$.
It is a pure and an entangled state. It describes the
equal superposition of single-excitation states.  For a chain with
$N+1$ elements this state is
\begin{equation}
|\Psi\rangle=\frac{1}{\sqrt{N+1}}\sum_{j=1}^{N+1}|1_j\rangle,
\label{stationary}
\end{equation}
where
\[|1_j\rangle=|+_j\rangle\prod\limits_{k\neq j}|-_k\rangle.\]
So, the state (\ref{stationary}) describes just one excitation "spread" homogeneously over all the systems of the chain.

Eqs. (\ref{chain2}) and (\ref{cohr1}) can be easily solved
analytically \cite{lawler}. For example, the solution for the
elements $\rho_{kl}(t)$ can be written as
\begin{equation}
\rho_{kl}(t)=\sum\limits_{m,n=1}^{N+1}\alpha^{kl}_{mn}\exp\{-\lambda_{m,n}t\},
\label{discrete solution1}
\end{equation}
where the eigenvalues for Eq. (\ref{chain2}) are
\[\lambda_{m,n}=4\gamma\left\{\sin^2\left(\frac{\pi m}{N+1}\right)+\sin^2\left(\frac{\pi n}{N+1}\right)\right\}.
\]
The coefficients $\alpha^{kl}_{mn}$ are defined from the initial state:
\[\alpha^{kl}_{mn}={\nu_{km}\nu_{ln}}\langle \Phi_m|\rho(0)|\Phi_n\rangle,
\]
where the coefficients are
\[\nu_{km}=\sqrt{(2-\delta_{k,N+1})}\cos\left(\frac{2\pi k}{N+1}m\right).
\]
The symbol $\delta_{ij}$ denotes Kronecker delta, and the vectors are
\begin{equation}
|\Phi_l\rangle=\frac{1}{\sqrt{N+1}}\sum\limits_{n=1}^{N+1}\nu_{ln}|1_n\rangle.
\label{discrete vectors}
\end{equation}
Vectors (\ref{discrete vectors}) are mutually orthogonal, $\langle \Phi_m|\Phi_n\rangle=\delta_{mn}$. However, they do not diagonalize the density matrix, $\rho$. Since $\lambda_{N+1,N+1}=0$, the vector $|\Phi_{N+1}\rangle$ is the stationary state $|\Psi\rangle$, see Eq. (\ref{stationary}).

Solution of Eq.(\ref{cohr1}) is similar to the solution of Eq. (\ref{chain2}).  Eq.~(\ref{chain2}) can also be viewed as a
discretization of the standard heat-transfer equation in the square $L\times L$, where $L=Na/2$ is the chain length, and $a$ is the doubled distance between neighboring systems in the original chain (the one depicted above in Fig.(\ref{fig1})).
The later one can be obtained by assuming $ak\rightarrow x$, $al\rightarrow
y$:
\begin{eqnarray}
\frac{d}{dt}\rho(x,y;t)\approx
a^2\gamma\left(\frac{\partial^2\rho(x,y;t)}{\partial x^2}
+\frac{\partial^2\rho(x,y;t)}{\partial y^2}\right).
 \label{heat1}
\end{eqnarray}
 The absence of quantum flux over the chain boundaries corresponds to the Neumann boundary conditions for the density matrix at the square boundaries $x=0,L$  and $y=0,L$. The discrete solution (\ref{discrete solution1}) approximate the solution for the Neumann boundary conditions only in the limit $N\rightarrow \infty$. For initial states having only non-negative elements,
$\rho_{kl}(t=0)\geq 0$, Eq.~(\ref{chain2}) gives
$\rho_{kl}(t>0)\geq 0$. Thus, elements $\rho_{kl}$ can be taken for
classical probabilities, and our dissipatively coupled 1D chain
can indeed simulate 2D classical walk or heat transfer. In Fig.~\ref{fig2}(a) examples of dynamics are shown
for the case of just one TLS being initially completely excited. Even for modest number of TLS in the chain ($N=51$ for our
example depicted in Fig.~(\ref{fig2})) solutions of
Eq.~(\ref{chain2}) given by Eq.(\ref{discrete solution1}) are very close to the fundamental solution of Eq. (\ref{heat1}) obtained in the limit of infinitely-length chain
\begin{equation}
\rho_0(x,y;t)=
\frac{1}{4a^2\gamma t}\exp\left\{-[(x-x_0)^2+(y-y_0)^2]/2a^2\gamma
t\right\},
\label{heat solution1}
\end{equation}
for $\gamma t > 1$. Here $x_0,y_0$ denote the
position of the initial excitation.

A continuous approximation for Eq.(\ref{cohr1}) is the 1D heat-transfer equation,
\begin{equation}
\frac{d}{dt}\rho(z;t)\approx a^2\gamma\frac{\partial^2\rho(z;t)}{\partial z^2}.
\label{heat01}
\end{equation}
All the consideration given above can be extended for this case, too.
So, our dissipatively coupled 1D chain can simulate simultaneously both 1D and 2D classical random walk, or heat transfer.

Non-exponential decay in our chain is stipulated by
nonlocality of Lindblad operators  in Eqs. (\ref{chain2}) and (\ref{cohr1})). It is
interesting and instructive to compare the dynamics of population
decay in our chain with dynamics of collective spontaneous
emission of dense atomic cloud with dipole-dipole interaction
\cite{scully2009,scully2010}. The continuous limit of master
equation correspondent to the model considered in Ref.~
\cite{scully2009,scully2010} is given by Eq.(\ref{wave function4}) in the
 \ref{append_a}. Eq.(\ref{wave function4}) looks similar to
Eq.~(\ref{heat1}) and also manifests an algebraic law of
population decay, $1/\sqrt{\gamma t}$ \cite{scully2009}. However,
it holds in 3D-space, and describes quite different physical
process. The reason for it is the formal equivalence of Eq.(\ref{wave function4})
to the master equation with nonlocal Lindblad operators. This
nonlocality arises from non-conservation of the excitation number
and from accounting for quantum states corresponding to two
excited atoms and one virtual photon with "negative" energy
\cite{scully2009}.

\section{Anomalous heat transfer}
\label{therm}

Eqs. (\ref{chain2}) and (\ref{cohr1})) tell us that elements of the density matrix in the energy eigenstates basis evolve according to classical equations.  However, despite this fact, the energy flow through the 1D chain cannot be described by the 1D classical Fourier law even in the limit of the large number of TLS in the chain. This holds even in the case, where the density matrix elements of the initial single-excitation state are non-negative.
Indeed, as it was shown in the previous Section (see Eqs. (\ref{conservation})),   the
sum of matrix elements,
\[W=\sum\limits_{k,l=1}^{N+1}\rho_{kl},\]
is preserved. Simultaneously, the total
upper-state population of the chain does decay. As it follows from Eqs. (\ref{chain2}), for example, for the initial state being just one completely excited TLS, the total population behaves in the following way:
\[\left\langle\sum\limits_{k=1}^{N+1}|1_k\rangle\langle1_k|\right\rangle=\sum\limits_{k=1}^{N+1}\rho_{kk}\propto\frac{1}{\sqrt{\gamma
t}}.
\]
A physical reason for breaking the
classical Fourier heat conductivity law
is rather simple. Besides the energy transport occurring in one
dimension (along the chain), there is an additional motion.
Fourier law is always breaking down in presence of additional
motions (the simplest example is the convection in liquids
\cite{landau}). In our case this additional motion is a flow of
quantum coherence. Indeed, in the continuous approximation the energy flux along the chain reads
as
\begin{equation}
J(x,x)\propto \lim\limits_{y\rightarrow
x}\frac{\partial}{\partial x} \rho(x,y;t).
\label{tr}
\end{equation}
The right-hand part of
this equation cannot be represented in the form of the gradient of
a scalar function satisfying the 1D heat-transfer
equation. Thus, the right-hand part of
Eq.(\ref{tr}), or the energy per TLS,  $\rho(x,x;t)$, cannot be associated with classical
temperature \cite{landau}.
Since in our chain not only  energy  is transported, but  also quantum coherences, it is
possible to introduce the concept of the effective 2D "quantum
temperature" and 2D quantum flux as, correspondingly,
\begin{eqnarray}
\nonumber
T_q(x,y;t)=\rho(x,y;t), \\
\nonumber
\vec{J}_q(x,y;t)\propto\nabla_{xy}\rho(x,y;t).
\end{eqnarray}
The value of the real energy flux corresponds to the
$x$-component of quantum flux, $\vec{e}_x\cdot\vec{J}_q(x,x;t)$,
where $\vec{e}_x$ is the unit vector along the chain, while the
$y$-component describes the coherences transfer. This additional
motion produces a second dimension in the heat transfer equation
and leads to the anomalous thermodynamics. A number of examples of an anomalous
thermodynamical behavior for quantum structures is known (see, for
example, Refs.\cite{volz,dubi,celani}). But the dissipatively coupled
chain differs starkly by the nature of anomaly. One can retrieve classical
thermodynamics  and establish connection between "classical" and "quantum" temperatures
by neglecting the additional motions in
Eq.(\ref{heat1}), i.e. by averaging out some coherences.  Indeed, introducing the averaged
temperature and averaged energy flux as
\begin{eqnarray}
\label{class1} T_{class}(x;t)=\int \rho(x,y;t)dy, \\
\vec{J}_{class}(x;t)=\int\vec{J}_q(x,y;t)dy \label{class2}
\end{eqnarray}
one obtains 1D heat transfer equation for temperature and relation
between the temperature and energy flux,
\[\vec{J}_{\rm class}(x;t)=\vec{e}_x\frac{\partial}{\partial x}
T_{\rm class}(x;t),\]
in full agreement with Fourier heat conductivity
theory. But there is no convincing reasons for such approximation to be done
and for the association of the integrals (\ref{class1}, \ref{class2}) with
a real physical temperature and a real energy flux. As it will be
shown below, use of such classical model in our case leads to the
disappearance of some important observable regimes.

\begin{figure}[htb]

\includegraphics[width=\textwidth]{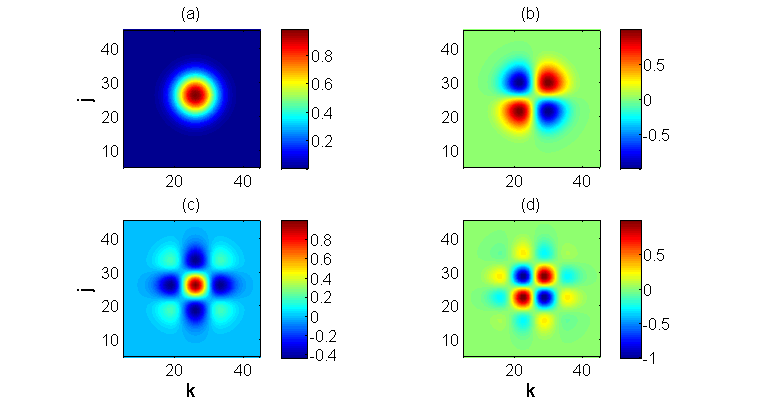}

\caption{ Contour plots of exact solutions of the master equation
Eq.~(\ref{chain2}) for initial states (\ref{initial general}). For
the panel (a) the initial state is the just one completely excited system, $m=0$. For the panel
(b) $m=1$, $f_0=f_1=1/\sqrt{2}$. For the panel (c) $m=2$,
$f_0=f_2=1/\sqrt{6}$, $f_1=2/\sqrt{6}$. For the panel (d) $m=3$,
$f_0=f_3=1/\sqrt{20}$, $f_1=f_2=3/\sqrt{20}$. For all the panels
$\gamma t=9$, $N=51$. For comparison, solutions are normalized (the largest element is taken to be unity).}
\label{fig3}
\end{figure}

\section{Polynomial decay of populations}
\label{polynomial}

The dissipatively  coupled chain offers unique possibility of controlling the quantum state dynamics.
Just by choosing different initial state, one can manipulate the law of  spontaneous decay. Also, dependence of the decay law on the type of initial state gives one an opportunity of identifying initial states of
the chain by  measuring the population of just one TLS.

We have obtained that for the
simple 1D set of TLS coupled to common Markovian reservoirs and
the single initially excited TLS, the upper-state population decay
for chain TLS occurs polynomially, according the the $1/t$ law.
Let us now demonstrate  by just adjusting the initial state
of the chain, one can obtain  a population decay law $1/t^{2m+1}$
for an arbitrary $m\ge0$. Interestingly, we can arrange initial conditions that are hardly possible in the classical case.
In particular, it is possible to create equivalents of neighboring
regions with "positive" and "negative" temperatures (or even
"imaginary" ones). It can be done by choosing an initial state
with components orthogonal to the stationary state (\ref{stationary}). An example of
$\rho_{kl}$ dynamics for such a state is demonstrated in
Fig.~\ref{fig2}(c), where the solutions of Eq.~(\ref{chain2}) are shown for entangled initial states:
\begin{equation}\rho_{\pm}=|\psi_{\pm}\rangle\langle\psi_{\pm}|,
\quad
|\psi_{\pm}\rangle=\frac{1}{\sqrt{2}}(|1_j\rangle\pm|1_{j+1}\rangle).
\label{initial entangled}
\end{equation}
The solution for the state $\rho_+$ behaves itself "classically":
both closely situated initial peaks are soon merged together in
one Gaussian shape. However, dynamics with the initial state
$\rho_{-}$ is drastically different: negative elements,
$\rho_{kl}$, do persist (Fig.~\ref{fig2} (d)), and  initially close
peaks are distancing from each other with time.
Moreover, excitation displacement for the initial state $\rho_{-}$
appears to be faster than for the initial state $\rho_+$. Such a
difference can be quite simply illustrated with the solution of the
continuous approximation (\ref{heat1}). The continuous analog of
$\rho_{\pm}$ can be represented as
\begin{eqnarray}
\nonumber
\rho_{\pm}(x,y;t=0)\propto\delta(x-x_0)\delta(y-y_0)+\delta(x-(x_0+a))\delta(y-(y_0+a)) \\
\nonumber
\qquad \pm\delta(x-(x_0+a))\delta(y-y_0)\pm\delta(x-x_0)\delta(y-(y_0+a)),
\end{eqnarray}
where $\delta(x)$ is the delta-function. It is easy to see that when
the excitation has spread for distances $x$, $y$ such that
\[\sqrt{(x-x_0)^2+(y-y_0)^2}\gg a,
\]
the solution for the initial
state $\rho_+$  is given by Eq.~(\ref{heat solution1}). However, for
the initial state $\rho_{-}$  dynamics is quite different. When
the excitation has spread far enough, the solution for the initial
state $\rho_-$ is
\[\rho(x,y;t)\approx2\rho_0(x,y;t){(x-x_0)(y-y_0)}/{(2a\gamma
t)^3}.\]
Thus, just by choosing the initial state to be $\rho_-$, we
get the population decay law $1/t^3$. This state doesn't exist within the bounds of quasiclassical
Fourier theory. Indeed, the averaging of this state following
(\ref{class1}, \ref{class2}) leads to the trivial solution: the
temperature as well as energy flux equal to zero at arbitrary
point of space for every moment of time. It is not hard to see
that the population decay law can be further varied by choice of
the initial state. Let us consider pure initial states of the
general form
\begin{equation}
|\phi\rangle=\sum\limits_{i=0}^{m}(-1)^i f_{i}|1_{k+i}\rangle,
\label{initial general}
\end{equation}
where $f_i$  are some scalar coefficients, and the excitation is
assumed to be well localized, $m\ll N$. Taking the continuous
approximation and using  the quantity $m/\gamma
t \ll 1$ as small parameter, from Eq.~(\ref{initial general})
we straightforwardly obtain  the following approximation:
\begin{eqnarray}
\nonumber
\rho(x,y;t)\approx\rho_0(x,y;t)\sum\limits_{n=0}^{\infty}\frac{1}{n!(2
a^2 \gamma t)^{n}} \\
\times\sum\limits_{i,j=0}^{m}(-1)^{i+j}f_{i}f_{j}[i(x-x_{0})+j(y-y_{0})]^{n}.
\label{approximation}
\end{eqnarray}
Starting with the initial state (\ref{initial general})
and choosing the coefficients $f_i$ in the form of the normalized binomial
coefficients, $f_{i}\propto{m!}/{i!(m-i)!}$, we get
the first
non-zero term in the approximation (\ref{approximation}) of
the order of $1/t^{2m+1}$. Examples of exact solutions for initial
states (\ref{initial general}) are shown in Figs.~\ref{fig3}. The panels (a-d) corresponds to $m=0,1,2,3$.

Notice, that the states (\ref{initial general}) are
non-correctly described within the bounds of the quasiclassical
Fourier theory. The "classical" averaging of states (\ref{initial
general}) as it is done in Eqs.~(\ref{class1},~\ref{class2}) can
lead to the drastic deformation of them (for example, the states
orthogonal to the stationary state will be reduced to zero by such
averaging). Mind that polynomial decay regimes considered here are
taking place for times, when initially localized excitation has
spread far enough. However, we do not consider here an influence
of edges. That is, we limit ourselves to the time intervals satisfying
$am<v(t)< aN$, where $v(t)$ is the variance of the distribution
$\rho_{kl}(t)$. For example, for the single initially excited TLS
this condition reads as $1<\gamma t <N$. An influence of chain
edges will be considered in further work.

\section{Multi-excitation dynamics}
\label{multidy}

It is easy to surmise from Eq.(\ref{chain1}) that there is a profound difference between chain dynamics in cases of one and multiple initial excitations. Indeed, the state with multiple excitations will inevitable decay toward mixture of the stationary state (\ref{stationary}) and the vacuum state. Generally, for the initial state with no more than $N$ excitation, equations of density elements corresponding to the $N$-excitation subspace describe $2N$-dimensional classical random walk with  losses  (see equations for an arbitrary number of initial excitations given in \ref{append_b}). However, even in this case one can still model lossless multi-dimensional random walks within some limited period of time and for some specific classes of initial states. Moreover, there are regimes when one can have preservation of coherences, where the chain behaves itself in quite "thermodynamic" way.

Let us illustrate our consideration with the case of no more than two initial excitations in the chain . Two-excitation matrix elements are
\[\rho^{k,l}_{m,n}=\langle 1_k,1_l|\rho|1_m,1_n\rangle,\]
where
\[|1_k,1_l\rangle=|+_k\rangle|+_l\rangle\prod\limits_{j\neq k,l}|-_j\rangle.\]
In difference with Eqs.(\ref{chain2}), there are two different kinds of equations for the matrix elements $\rho^{k,l}_{m,n}$. For the elements without neighbouring indexes, i.e.  for $k\neq l\pm1$ and $m\neq n\pm1$,
from Eq.(\ref{chain1}) one gets the following system of equations (for the sake of illustration we are giving here equations only for the internal TLS of the chain, i.e. $k,l,m,n\neq 1,N+1$):
\begin{eqnarray}
\nonumber
\frac{1}{\gamma}\frac{d}{dt}\rho^{k,l}_{m,n}=-8\rho^{k,l}_{m,n}+\rho^{k+1,l}_{m,n}+\rho^{k-1,l}_{m,n}\\
\label{chain4}
\qquad +\rho^{k,l-1}_{m,n}+\rho^{k,l+1}_{m,n}+\rho^{k,l}_{m-1,n}\\
\nonumber
\qquad +\rho^{k,l}_{m+1,n}+\rho^{k,l}_{m,n-1}+\rho^{k,l}_{m,n+1}.
\end{eqnarray}
Obviously,  Eq.(\ref{chain4}) coincides with the equation for 4D classical random walk and in the continuous limit transforms to the 4D heat-transfer equation.

The situation is quite different for the matrix elements with neighbouring indexes. Let us assume, for example, that $l=k+1$. Then, instead of Eqs.(\ref{chain4}) one has
\begin{eqnarray}
\nonumber
\frac{1}{\gamma}\frac{d}{dt}\rho^{k,k+1}_{m,n}=-8\rho^{k,k+1}_{m,n}+\rho^{k-1,k+1}_{m,n}+\rho^{k,k+2}_{m,n} \\
\label{chain5}
\qquad +\rho^{k,l}_{m-1,n}+\rho^{k,l}_{m+1,n}+\rho^{k,l}_{m,n-1}+\rho^{k,l}_{m,n+1}.
\end{eqnarray}
Eq.(\ref{chain5}) does not coincide with the equation for classical random walk. It contains terms describing loss. It can be easily seen in the continuous limit, since then, for example,  one has
\[\rho^{k,k+2}_{m,n}-\rho^{k,k+1}_{m,n}\rightarrow \lim\limits_{x_2\rightarrow x_1}a\frac{\partial}{\partial x_2}\rho(x_1,x_2,y,z;t),
\]
where variables $x_1,x_2,y,z$ correspond to the indexes $k,l,m,n$, respectively. Of course, it should be expected. Multi-excitation state does decay toward the single-excitation one.

Eqs. (\ref{chain4}) and (\ref{chain5}) point to a number of quite counter-intuitive conclusions. First of all, if the initially excited TLS are far from each other, the spread of coherences occurs as for lossless multi-dimensional random walk  till the excitation spreads to neighbouring TLS. Notice, that it will take place even in the case of entangled initial state of several TLS. If one has initially several uncorrelated excited TLS being far from another, the whole chain behaves like a set of unconnected chains with just one excitation per chain till the excitation spreads to neighbouring TLS. So, it means that one might have a sort of conservation of coherences in this case, too.

Let us demonstrate an appearance of such a "conservation law of a sorts" with the example of two initially non-interacting chains with no more that one excitation in each. So, we take that there are chains of $M$ and $N$ TLS with both $M,N\gg1$. Also, corresponding sums of coherences (\ref{conservation}), denoted as $W_M$ and $W_N$ are supposed to be much less than number of TLS, $W_X\ll X$, $X=M,N$. We assume that the chains are initially in the stationary states,
\begin{eqnarray}
\rho_{X}^{st}=\frac{W_{X}}{X}|\Psi_{X}\rangle\langle\Psi_{X}|+
\left(1-\frac{W_{X}}{X}\right)|V_X\rangle\langle V_X|,
\label{spreader2}
\end{eqnarray}
where $X=M,N$, the vectors $|\Psi_{X}\rangle$ are given by Eq.(\ref{stationary}) and vectors $|V_X\rangle$ describe the lowest energy state of corresponding chains.

Now let us assume that the chains are coupled by the common dissipative reservoir connecting the last TLS of the first chain and the first TLS of the second chain (the rate of decay to this reservoir we take to be the same $\gamma$ as the rates for all other reservoirs). Then, the stationary state of compound chain will be obviously given also by Eq.(\ref{spreader2}) with $X=M+N$. It is easy to get from Eqs.(\ref{chain4}) and (\ref{chain5}) that the sum of coherences for the new stationary state is
\begin{equation}
W_{N+M}=W_N+W_M+O\left(W_NW_M\left(\frac{1}{M}+\frac{1}{N}\right)\right)\approx W_N+W_M.
\label{WMN}
\end{equation}
 Notice that the expression (\ref{WMN}) becomes exact in the limit $M,N\rightarrow\infty$. So, the sum of coherences for the stationary state of the compound chain is indeed approximately equal to the same of coherences of parts. Moreover, if one disconnects chains and they settle into stationary states again, the value of matrix elements of single-excitation density matrix (the "quantum temperature" as introduced in the Section \ref{therm}) for both part will remain to be equal, $\rho^{X}_{kl}=(W_{M+N}/(M+N)^2)$.

\section{Conclusions}
We have suggested and discussed a tight-binding dissipatively
coupled quantum chain. It consists of two-level systems pairwise
coupled to the same Markovian reservoirs. We have shown that
despite being composed of a comparatively few systems forming the
simplest one-dimensional chain, such a chain can model a
multi-dimensional random walk, or model a solution of
multi-dimensional heat-transfer equation. In difference to the
classical "original", it is possible to model heat flow from
initial distributions with regions of positive and negative
temperatures. Such possibility of quantum stimulations in
many-body physics corresponds to the long-standing idea first
proposed by R. Feinman and recently cited by C. Cohen-Tannoudji
and D. Guery-Odelin as a conclusive remark to their book
\cite{cohen}. This model exhibits anomalous thermodynamics
behavior, such as non-Fourier heat conductivity and non-existence
of temperature in the classical meaning. The population dynamics
of all TLS in the chain is always non-exponential and exhibit
polynomial character. By the choice of the initial state,
different power laws of decay, $1/t^{2m+1}$, for an arbitrary
$m>0$ can be achieved. The suggested chain can be used as an
efficient simulator of  classically hard problems, such as
multi-dimensional quantum walks or heat-transfer equations. Note,
that the considered chain is not unique. One can build a number of
different dissipatively coupled chains, for example, by changing
phases in the system-system interaction terms in the chain.

G. S. acknowledges support from the EU FP7 projects FP7 People
2009 IRSES 247007 CACOMEL and FP7 People 2013 IRSES 612285 CANTOR.
This work was also supported by the National Academy of Sciences
of Belarus through the program "Convergence", by the External
Fellowship Program of the Russian Quantum Center at Skolkovo (D.
M., S. K. and E. G.) and by FAPESP grant  2014/21188-0 (D. M.), N.
K. and D. M. acknowledge the support provided by the Scottish
Universities Physics Alliance (SUPA). The research leading to
these results has received funding from the European Community
Seventh Framework Programme (FP7/2007-2013) under grant agreement
n.~270843 (iQIT). Authors are thankful to  Dr. B. W. Lovett, Dr.
Sh. Starobinets and Prof. Miguel A. Martin-Delgado for helpful
discussions and pointing to relevant references.

\newpage
\appendix
\section{Comparison of dissipatevely coupled quantum chain with spontaneously emitted atomic cloud}
\label{append_a}

It is interesting and constructive to compare single-excitation
dynamics of the dissipatively coupled quantum chain with
collective spontaneous emission in the dense atomic cloud with
dipole-dipole interactions (see Ref.\cite{scully2010}). In
\cite{scully2010} in the main text a system of $N$ TLS is considered.
Initially, one of them is in the excited state, while all others
are in the ground state. TLS are placed at positions $\vec{r}_j$,
$j=1,2\ldots N$. Multi-mode electromagnetic field is interacting
with all TLS. Initially, this field is in the vacuum state,
$|vac\rangle$. The solution of the Schrodinger equation for the
atoms and field can be represented as
\begin{eqnarray}
|\Psi(t)\rangle=\sum\limits_{j=1}^N\beta_j(t)|1_j\rangle|vac\rangle+|\Psi_{rest}(t)\rangle,
\label{wave function1}
\end{eqnarray}
where the coefficients $\beta_j(t)$ are amplitudes of having $j$th
TLS completely excited; the wave function $|\Psi_{rest}(t)\rangle$
denote components of the total wave function, $|\Psi(t)\rangle$,
having other then single excited TLS and the field vacuum. Notice
that in Ref. \cite{scully2010} of the main text the rotating-wave approximation
was not used to describe TLS-field interaction. So, the function
$|\Psi_{rest}(t)\rangle$ includes also components with the number
of excitations larger than one.

Using standard approximations about the reservoir, in Ref. \cite{scully2010} the following system of equations was obtained
\begin{eqnarray}
\frac{d}{dt}\beta_j(t)=-\gamma\beta_j(t)
+i\gamma\sum\limits_{k\neq j}\beta_k(t)K_{jk}, \label{wave function2} \\
K_{jk}=\frac{\exp\{-ik_0|\vec{r}_k-\vec{r}_j|\}}{k_0|\vec{r}_k-\vec{r}_j|},
\label{matrix}
\end{eqnarray}
where $k_0=\omega_0/c$, $\omega_0$ is the TLS transition
frequency, $\gamma$ is the decay rate into the field reservoir.

Eq.(\ref{wave function2}) can be easily transformed to the form
similar to Eq.(\ref{chain2}) of the main text. Indeed, introducing the matrix
elements $\rho_{jk}=\beta_j\beta_k^*$, from Eq. (\ref{wave
function2}) it follows that
\begin{eqnarray}
\nonumber
 \frac{d}{dt}\rho_{jk}=-2\gamma\rho_{jk}
+i\gamma\left(\sum\limits_{l\neq
j}\rho_{jl}K_{jl}-\sum\limits_{l\neq k}\rho_{lk}K_{lk}^*\right).
\label{wave function3}
\end{eqnarray}
Continuous analog of Eq. (\ref{wave function3}) reads
\begin{eqnarray}
\nonumber
\frac{d}{dt}\rho(\vec{r},\grave{\vec{r}})=-2\gamma\rho(\vec{r},\grave{\vec{r}})\\
\qquad +i\gamma\frac{N}{V}\int\limits_V
d^3\vec{R}\frac{\exp\{-ik_0|\vec{r}-\vec{R}|\}}{k_0|\vec{r}-\vec{R}|}\rho(\vec{R},\grave{\vec{r}})\\
\nonumber
\qquad -i\gamma\frac{N}{V}\int\limits_V
d^3\vec{R}\frac{\exp\{ik_0|\grave{\vec{r}}-\vec{R}|\}}{k_0|\grave{\vec{r}}-\vec{R}|}\rho(\vec{r},\vec{R}),
\label{wave function4}
\end{eqnarray}
where $V$ is the volume of the cloud. Equations Eq.(\ref{wave
function3}), Eq.(\ref{wave function4}) are similar to the Eqs. (\ref{chain2}),
(\ref{approximation}), respectively. The difference consists in the
form of operators in right-hand parts (integral operators with the
Green function of Helmholz equation in the capacity of kernel in
Eq.(\ref{wave function4}) instead of Laplace operator in Eq.(\ref{approximation}).  Integral operators in Eq.(\ref{wave function4})
are non-Hermitian due to the presence of counter-rotating terms in
Eq.(\ref{wave function2}) which are stipulated by the virtual
photons with "negative" energy. It lead to the matrix elements
(\ref{matrix}) instead of
\begin{eqnarray}
K_{jk}=i\frac{\sin\{k_0|\vec{r}_k-\vec{r}_j|\}}{k_0|\vec{r}_k-\vec{r}_j|}
\label{matrix1}
\end{eqnarray}
As a result, Eq (\ref{wave function3}) is not of Liouville type.
It is equal (as Eq. (\ref{approximation}) in the main text) to the master equation
with non-local Linblad operators. However,  it describes rather
different physical process. Thus, as it was shown in Ref.\cite{scully2010} in
the main text, collective spontaneous emission manifests for some
types of boundary conditions algebraic law of population decay.
Such behavior is stipulated by the virtual photons with "negative"
energy and disappears in the rotation-wave approximation (done by
replacing Eq. (\ref{matrix}) with Eq.(\ref{matrix1})).

\section{Equations for the matrix elements for the case of multiple excitations}

\label{append_b}

Here we write down equations for  matrix elements corresponding to states with the highest possible number of excitations.
Let us assume that initially we have no more than $m$ excitations
in the chain (for example, $m$ TLS in completely excited state).
From Eq.(\ref{chain1}) it is not hard to obtain the following set
of equations for the matrix elements corresponding to
$m$-excitation subspace
\begin{eqnarray}
\frac{d}{d t} \langle m_{k_{1},k_{2},\ldots,k_{m}}|\rho|m_{l_{1},l_{2},\ldots,l_{m}}\rangle=\sum_{i=1}^{N+1}\lambda_{i}+\mu_{i},
\label{multi}
\end{eqnarray}
where
\begin{eqnarray}
\nonumber \lambda_{i}=\{\gamma_{k_{i}}(\langle m_{\{k_{j}+\delta_{j,i}\}}|-\langle m_{\{k_{j}\}}|) \\
\qquad +\gamma_{k_{i-1}}(\langle m_{\{k_{j}-\delta_{j,i}\}}|-\langle m_{\{k_{j}\}}|)\}\rho|m_{\{l_i\}}\rangle, \\
\nonumber
\mu_{i}=\langle m_{\{ k_{i}\}}|\rho\{\gamma_{l_{i}}(|m_{\{l_{j}+\delta_{j,i}\}}\rangle-|m_{\{l_{j}\}}\rangle)\\
\nonumber
\qquad +\gamma_{l_{i}-1}(|m_{\{l_{j}-\delta_{j,i}\}}\rangle-|m_{\{l_{j}\}}\rangle)\}.
\label{lambda}
\end{eqnarray}

Here the state
$|m_{\{l_{j}\}}\rangle=|m_{l_{1},l_{2},\ldots,l_{m}}\rangle$ is
the state with $m$ TLS with numbers $l_{1},l_{2},\ldots,l_{m}$
being completely excited. Also,
$|m_{\{l_{j}\pm\delta_{j,i}\}}\rangle=|m_{l_1,l_2,\ldots,l_{i}\pm1,\ldots,l_{m}}\rangle$.
As it is in Eq.(3) of the main body of the paper, we are assuming
that $\gamma_{0}\equiv0$, and $\gamma_{k>0}\equiv\gamma$. Notice
that in case when the indexes are neighbors, i.e. when
$k_{i+1}=k_{i}+1$, one has
\begin{eqnarray}
\nonumber
\lambda_{i}=-\gamma_{k_{i}}\langle m_{\{k_{j}\}}|\rho |m_{\{l_i\}}\rangle+\\
\nonumber
\qquad \gamma_{k_{i-1}}\langle m_{\{k_{j}-\delta_{j,i}\}}|\rho |m_{\{l_i\}}\rangle-\langle m_{\{k_{j}\}}|\rho |m_{\{l_i\}}\rangle, \\
\nonumber
\lambda_{i+1}=\gamma_{k_{i+1}}\langle m_{\{k_{j}+\delta_{j,i+1}\}}|\rho |m_{\{l_i\}}\rangle-\\
\nonumber
\qquad \gamma_{k_{i+1}}\langle m_{\{k_{j}\}}|\rho |m_{\{l_i\}}\rangle- \gamma_{k_{i}}\langle m_{\{k_{j}\}}|\rho |m_{\{l_i\}}\rangle.
\end{eqnarray}
The same holds for the case of neighboring $l_i$ and $l_{i+1}$.
For indexes equal to $1$ or $N+1$ corresponding $\lambda$s are
\begin{eqnarray}
\nonumber
\lambda_{1}=\gamma_{k_{1}}(\langle m_{k_{1}+1,k_{2},...,k_{m}}|-\langle m_{k_{1},k_{2},...,k_{m}}|)|\rho |m_{\{l_i\}}\rangle,\\
\nonumber
\lambda_{N+1}=\gamma_{k_{N}}(\langle m_{k_{1},k_{2},...,k_{m}-1}|-\langle m_{k_{1},k_{2},...,k_{m}}|)|\rho |m_{\{l_i\}}\rangle.
\end{eqnarray}

For matrix elements without
neighboring excitations, i.e. $k_i\neq k_j\pm1$,  $l_i\neq
l_j\pm1$, $\forall k_i,k_j$,
one obtains that Eq.(\ref{multi}) is the discretization of the
following $2m$-dimensional heat-transfer equation
\begin{eqnarray}
\nonumber \frac{d}{dt}\rho(\{x_k\},\{y_k\};t)\approx
a^2\gamma\sum_{k=1}^m\frac{\partial^2\rho(\{x_k\},\{y_k\};t)}{\partial x_k^2}
\\
\qquad \qquad +a^2\gamma\sum_{k=1}^m\frac{\partial^2\rho(\{x_k\},\{y_k\};t)}{\partial y_k^2}.
\label{heat4}
\end{eqnarray}

\end{document}